\documentclass[11pt]{article}
\usepackage{moriond,epsfig}

\def\be{\begin{equation}}
\def\ee{\end{equation}}
\def\bea{\begin{eqnarray}}
\def\eea{\end{eqnarray}}

\newcommand{\diff}[2]{\frac{\mathrm{d}#1}{\mathrm {d}#2}}

\begin{document}
\vspace*{4cm}
\title{INDIRECT DARK MATTER SEARCH: COSMIC POSITRON FRACTION MEASUREMENT FROM 1~TO~50~GEV WITH AMS-01}

\author{H.~GAST, J.~OLZEM, and ST.~SCHAEL}

\address{I.~Physikalisches Institut, RWTH Aachen,\\
Sommerfeldstra\ss{}e 14, 52074 Aachen, Germany}

\maketitle\abstracts{
A new measurement of the cosmic ray positron fraction $e^+/(e^++e^-)$
in the energy range of 1-50~GeV is presented. The measurement is
based on data taken by the AMS-01 experiment during its 10 day space
shuttle flight in June 1998. A proton background suppression in the
order of $10^{6}$ is reached by identifying converted bremsstrahlung
photons emitted from electrons and positrons.}

\section{Introduction}
Over the past decades, cosmic ray physics has joined astronomy as a
means to gather information about the surrounding universe. Of the few
particles that are stable and thus able to cross the vast interstellar
distances, electrons and positrons are of particular
interest.

Electrons are believed to be accelerated in shock waves following
supernova explosions. Their spectrum is subsequently altered by
inverse Compton-scattering off cosmic microwave background photons,
synchrotron radiation due to the galactic magnetic field,
bremsstrahlung processes in the interstellar medium, and finally,
modulation in the solar magnetosphere. Thus, they serve as an
important probe of cosmic ray propagation models. On the other hand,
positrons are secondarily produced in the decay cascades of $\pi^+$, created
in hadronic interactions of cosmic ray protons with the interstellar
medium, which yields an $e^+/e^-$ ratio of roughly $10\,\%$.

In addition to these classical sources, positrons may also originate
from more exotic ones. Among the most important problems in modern
cosmology is the nature of
dark matter. Based on observations of the cosmic microwave background,
supernovae of type IA, and galaxy clustering, among others, the
standard model of cosmology now contains a density of non-luminous
matter exceeding that of baryonic matter by almost a factor of
five \cite{eidelmann04}. The most promising candidate for dark matter
is a stable weakly interacting massive particle predicted by certain
supersymmetric extensions to the standard model of particle
physics \cite{turner90}, and called neutralino $\chi$. Positrons and
electrons will then be created in equal numbers as stable decay
products of particles stemming from $\chi$-$\chi$-annihilations, for
instance in the galactic halo. Such a process would constitute a
primary source of positrons. Therefore, a measurement of the positron
fraction is also motivated by the prospect of indirect dark matter
detection, especially if combined with other sources of information,
such as antiprotons, diffuse $\gamma$-rays or -- more challenging --
antideuterons.

\section{The AMS-01 experiment}
As a predecessor to the Alpha Magnetic Spectrometer AMS-02, which is
to be operated on the International Space Station (ISS) for at least 3
years, the AMS-01 experiment was flown on the Space Shuttle {\it Discovery}
from June 2nd to 10th, 1998.

The AMS-01 experiment is shown in fig.~\ref{fig:ams3D}. It consisted of
a cylindrical permanent magnet with a bending power of 0.14 Tm$^2$ and
an acceptance of 0.82 m$^2$sr. The magnet bore was covered at the
upper and lower end each with two orthogonal layers of scintillator
paddles, forming the time of flight system (TOF), which provided a
fast trigger signal as well as a measurement of velocity and charge
number. Mounted inside the magnet volume, the tracking device
consisted of six layers of double-sided silicon strip detectors with
an accuracy of 10 $\mu$m in the bending coordinate. The inner magnet
surface was lined with the scintillator panels of the anticoincidence
system (ACC) serving as a veto counter against particles traversing
the magnet wall. For velocity measurements, AMS-01 had a two-layered
aerogel threshold \v Cerenkov counter (ATC) mounted underneath the lower TOF
layer, allowing e$^{+}$/p discrimination below 3 GeV/c. A low energy
particle shield covered the experiment to absorb particles below
5 MeV/c, while a multi layer insulation blanket (MLI) served as a
protection against space debris and solar radiation. The radiation
thickness of all materials above or below the tracking device, not
including the Space Shuttle, sums up to 18.2\% or 19.1\% of a
radiation length, respectively. A detailed
description of the experiment is given elsewhere \cite{aguilar02a}.

\section{Conversion of bremsstrahlung photons}
The main challenge of cosmic ray positron measurements is the
suppression of the vast proton background. As it is known from
previous measurements \cite{aguilar02a} \cite{duvernois01a}, the flux of
cosmic ray protons exceeds that of positrons by a factor of $10^4$ in
the momentum range of 1 - 50 GeV/c. Hence, in order to keep the proton
contamination of positron samples below 1\%, a proton rejection of
$10^6$ has to be reached. Since the ATC subdetector of AMS-01 provided
a sufficient single track proton rejection only for energies below 3
GeV, a different approach has been chosen for this analysis. It relies
on the identification of bremsstrahlung emission through
photoconversion. Due to the inverse quadratical dependence on the
particle mass of the cross section, bremsstahlung emission is
suppressed by a factor of more than $3\cdot 10^{6}$ for protons with
respect to positrons.
\begin{figure}[!h]
\begin{center}
\begin{tabular}{cc}
\epsfig{file=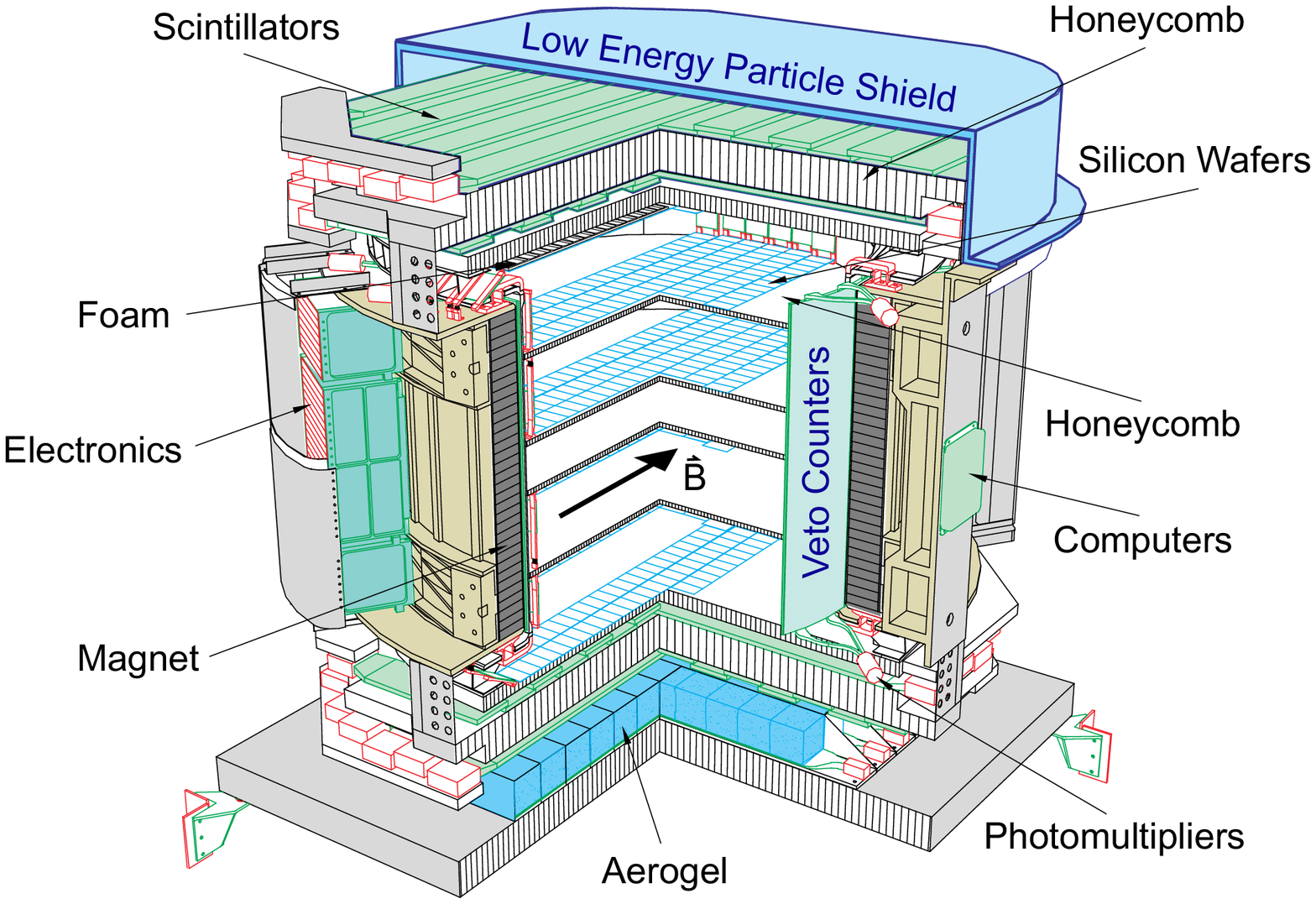,width=8.7cm}
&
\epsfig{file=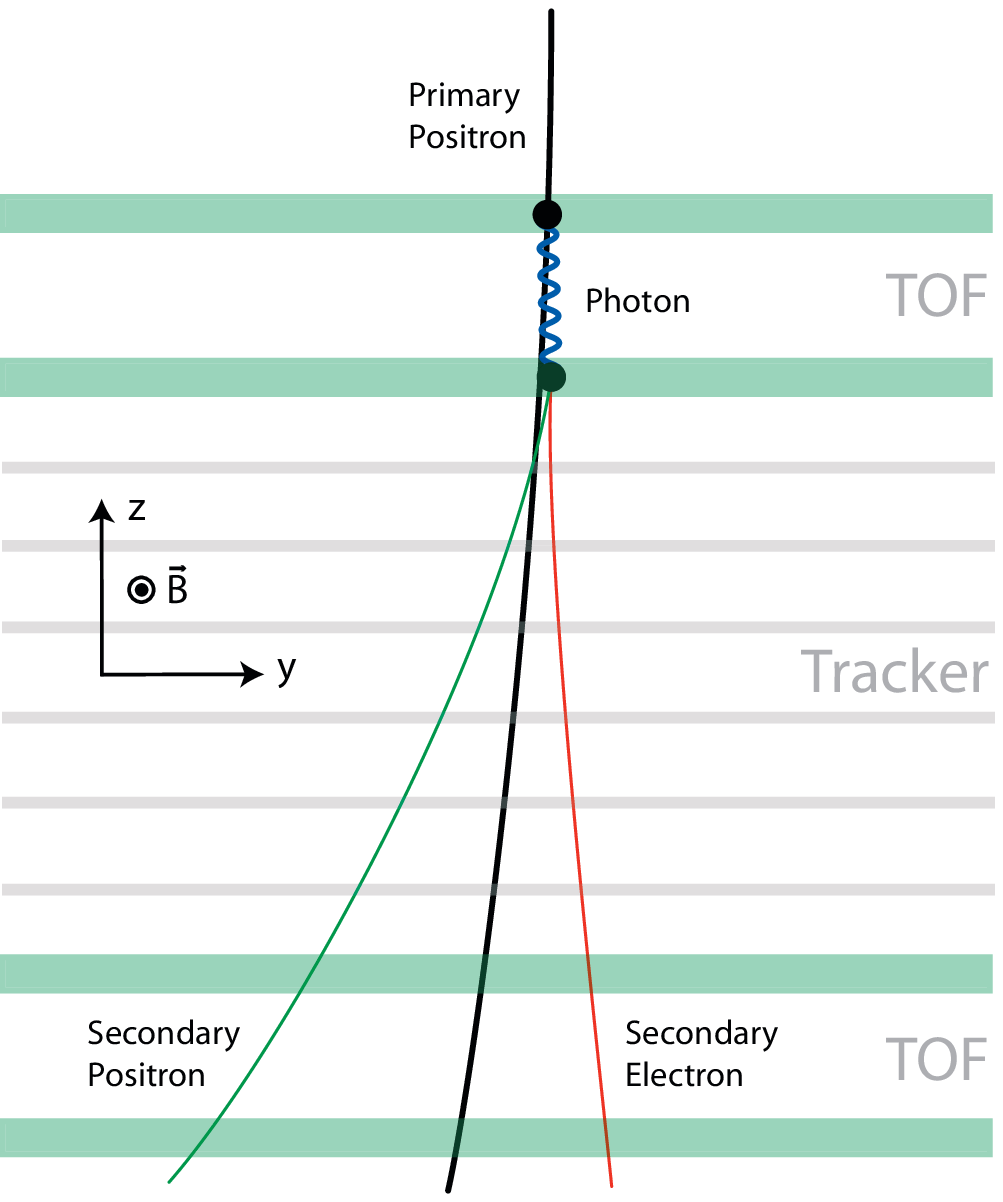,width=6cm}
\\
\end{tabular}
\end{center} 
\caption{\label{fig:ams3D}{\it Left: }The AMS-01 experiment. The tracker layers are
equipped asymmetrically. {\it Right: }
\label{fig:convScheme}
Schematic view of a converted
bremsstrahlung event caused by a positron going top-down.}
\end{figure}

Fig.~\ref{fig:convScheme} shows the principle of a converted
bremsstrahlung event signature. Here, a primary positron enters the
detector volume from above and emits a bremsstrahlung photon in
the first TOF scintillator layer. The photon then converts into an
electron positron pair, for example, in the second TOF layer. Because
of the low fraction of momentum which is typically carried away
by the photon, the secondary particles have lower momenta than the
primary.  Therefore, in the bending plane projection, the secondaries
tend to form the left and right tracks, while the primary remains in the
middle. 

Both bremsstrahlung and photon conversion are closely related
electromagnetic processes whose energy angle distributions can be
calculated with the Bethe-Heitler formalism. In the relativistic
limit, the angle of photon emission as well as the opening angle of
pair production show distributions with a most probable value of
$\theta_0\approx 1/\gamma$, $\gamma$ being the Lorentz factor of
the emitting particle or the electron positron pair, respectively.
In the GeV energy range, these values fall below the accuracy
limit of the track reconstruction induced by multiple scattering,
and thus are practically equal to zero.

The dominant background is caused by electrons with wrongly reconstructed
momentum sign, as well as by protons undergoing hadronic reactions in
the material distribution of the experiment. In the latter case mesons
are produced, which mimic the 3-track signature of converted
bremsstrahlung events. For example, in the reaction $pN\to
pN\pi^+\pi^-+X$, beneath additional undetected particles $X$, the
charged pions can be misidentified as an electron positron
pair. Besides this, neutral pions produced in reactions of the type
$pN\to pN\pi^0+X$ decay into two photons, one of which may escape
undetected.  If the remaining photon converts, the conversion pair
forms a 3-track event together with the primary proton.  However, the
invariant mass of the mesons and primary protons is typically at the
scale of the pion mass, leading to emission angles significantly
larger than zero.

\section{Analysis}
\label{section:analysis}
In order to gain the highest possible selection efficiency, it is
mandatory to apply sophisticated track and vertex finding algorithms
which are particularly customized for the signature of converted
bremsstrahlung \cite{jan} \cite{henning}.

Analysis and suppression of background mainly rely on the evaluation
of the topology and geometrical properties of the reconstructed
events, and are therefore based on data from the
tracker. Additionally, cuts on data from the TOF system are
applied. However, substantial parts of the analysis deal with measures
to account for the environmental circumstances under which the AMS-01
experiment was operated, especially the effect of the geomagnetic
field.

\subsection{Suppression of dominant background}
For the suppression of background, the fact is used that
bremsstrahlung and photon conversion imply small opening angles
of the particles at the vertices. In order to make these angles
independent of the frame of reference, the corresponding invariant
mass is calculated according to
\begin{equation}
m_{inv}^2 = 2 \cdot E_1 \cdot E_2 \cdot \left( 1 - \cos\theta \right),
\end{equation}
where $\theta$, $E_1$ and $E_2$ denote the opening angle and the
energies of the primary particle and the photon, or the conversion
pair, respectively.

\begin{figure}[h] 
\begin{center}
\begin{tabular}{cc}
\epsfig{file=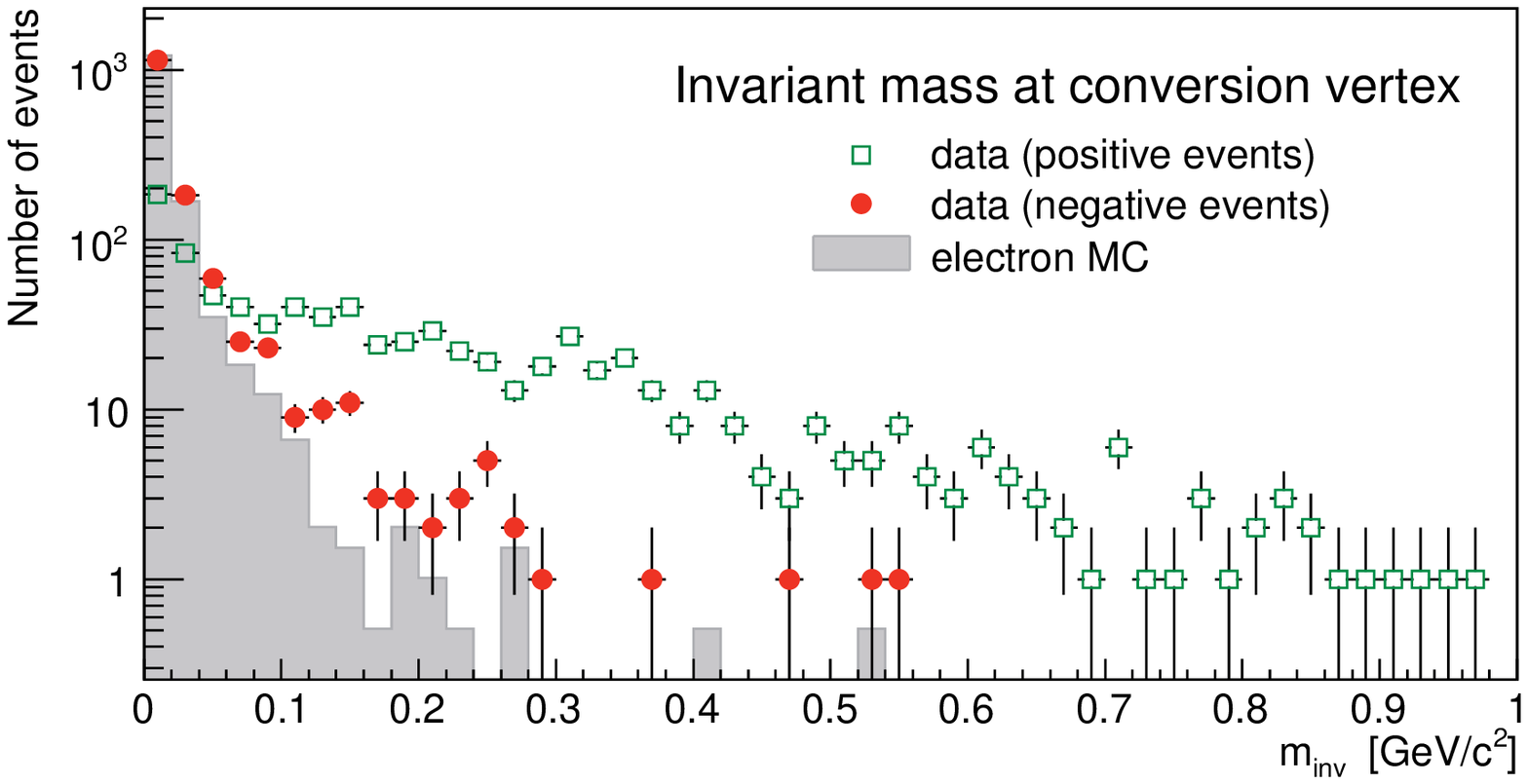,width=7.5cm}
&
\epsfig{file=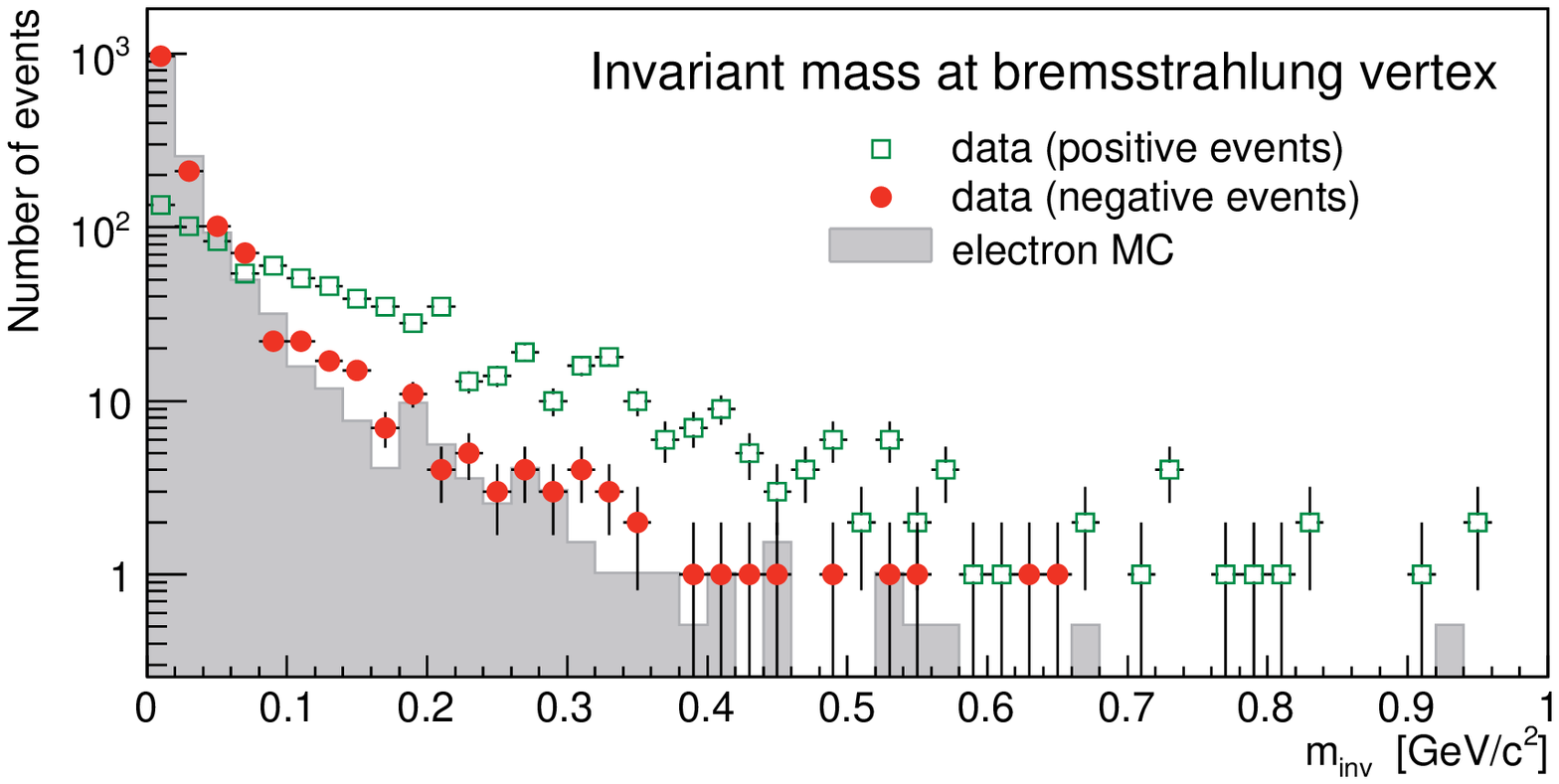,width=7.5cm}
\\
\end{tabular}
\caption{\label{fig:InvMass} The distribution of the
invariant mass at the conversion vertex {\it (left)} and at the 
bremsstrahlung vertex {\it (right)} for data {\it (dots)} and electron Monte Carlo {\it (shaded histogram)}.}
\end{center} 
\end{figure}

The distributions of the invariant masses are shown in
fig.\ref{fig:InvMass}.
For events with negative charge, which represent a largely clean
electron sample, they reveal a narrow shape with a peak at zero. This
is in perfect agreement with Monte Carlo results. In case of events with
positive total charge, consisting of positrons and background, the
distributions also show a peak at zero, but additionally a long tail
towards higher invariant masses caused by the proton background. 
In order to discriminate against background events, cuts are applied
on the invariant masses. The cuts are parameterized as ellipses in the
invariant mass plane.

The efficiency achieved reaches a flat maximum at approximately 20 GeV/c.
Towards higher momentum, the decreasing cluster separation approaches
the resolution limit of the silicon strip detectors. At low momentum, by contrast, secondary particles
may be deflected such that they generate multiple separated hits in
the TOF scintillators, which are rejected by the 
trigger logic of the experiment.

\subsection{Geomagnetic cutoff}
\label{subsection:cutoff}
Energy spectra of cosmic rays are modulated by the geomagnetic
field. Depending on the incident direction and the geomagnetic
coordinates of the entry point into the magnetosphere, particles with
momenta below a certain cutoff are deflected by the magnetic field and
cannot reach the Earth's proximity. Hence, below geomagnetic
cutoff the particles detected by AMS-01 must originate from within the
magnetosphere. They were mostly produced as secondaries through
hadronic interactions and trapped inside the Earth's radiation belts.

To discriminate against these secondaries, particle trajectories were
individually traced back from their measured incident location, angle
and momentum through the geomagnetic field by numerical integration of
the equation of motion \cite{flueckiger90a}. A particle was rejected
as a secondary if its trajectory once approached the surface of the
Earth, and thus originated from an interaction with the
atmosphere. Particles which did not reach a distance of 25 Earth radii
within a reasonable time were considered as trapped and also rejected.

\section{Correction for irreducible background}
\label{section:correction}
The distribution
of protons in the invariant mass plane does not vanish in the signal
region. The same applies to the background from misidentified
electrons. Consequently, a small fraction of background events will
not be rejected by the cut on the invariant masses. This remaining
irreducible background has to be corrected for. This has been
accomplished using Monte Carlo simulations.

The approach used is to run the analysis on an adequate number
of proton and electron Monte Carlo events as if they were data,
determine the amount and momentum distribution of particles that are
misidentified as positrons, and subtract these from the raw positron
counts obtained from data. However, such a comparison of
Monte Carlo and data requires the adjustment of several properties of
the simulated events: neither have they been affected by the
geomagnetic field, nor is their input spectrum exactly equal to the
true fluxes.

The influence of the geomagnetic field has been introduced to the
Monte Carlo particles by folding the distribution of measured momenta
with the normalized livetime function (see section \ref{section:fluxcalculation}). Deviations
of the incident momentum spectrum $\phi_{MC}(p)$ of Monte Carlo particles
from the true fluxes $\phi_D(p)$ measured by
AMS-01 \cite{aguilar02a} \cite{alcaraz00a} are corrected for by
reweighting the event variables with the ratio $\phi_D(p) /
\phi_{MC}(p)$.

\begin{figure}[!h]
\begin{center}
\epsfig{file=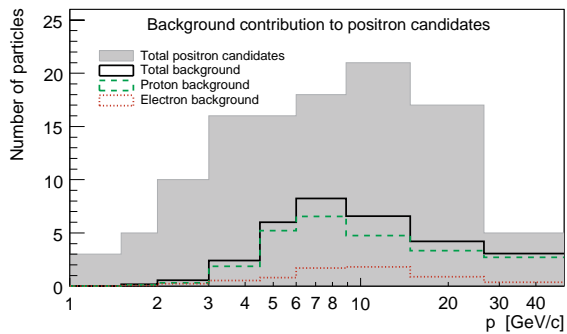,width=7.5cm}
\caption{\label{fig:subtractedBackground} Momentum distribution of the positron candidates 
including background, and the estimated background contribution from protons and
misidentified electrons.}
\end{center} 
\end{figure}
The number of proton Monte Carlo events is scaled to the data by using
the sidebands of the invariant mass distributions, while the electron
scaling can be determined directly from the candidate sample in the data.
Then, the background contribution to the number of positron candidates can be
calculated. Fig.~\ref{fig:subtractedBackground} shows the total
background correction as a function of momentum, separately indicating
the contributions from protons and misidentified electrons. They amount
to 24.9 and 6.4 events, respectively.

\section{Positron fraction}
\label{section:fraction}
The positron fraction $e^+/(e^{+}+e^-)$ is calculated from the
electron counts and corrected positron counts for each energy bin.  It
is shown in fig.~\ref{fig:fraction} in comparison with earlier results
and a model calculation based on purely secondary positron
production.

Due to the complexity of the positron fraction computation, taking
into account 2 sources of background, and low statistics, a Bayesian
approach based on Monte Carlo simulation has been chosen for the
determination of statistical errors.

In the positron fraction -- as a ratio of particle fluxes -- most sources
of systematic error, such as detector acceptance or trigger efficiency,
naturally cancel out. Only sources of error which are asymmetric
with respect to the particle charge have therefore to be considered.

The systematic error from background correction can be estimated by
evaluating the deviation of the scaled Monte Carlo background from the
data in the invariant mass plane. With a binning coarse enough to
flatten statistical fluctuations, the mean deviation outside the
signal region, i.e. $m_{inv} > 0.25$~GeV/c$^2$ at both vertices, leads
to a systematic error estimate of 20~\% of the background events. This
value is then propagated to the positron fraction for each momentum
bin.

As a consequence of the East West Effect, in combination with the
asymmetric layout of the AMS-01 tracker, the product of the detector
acceptance and the livetime as functions of the particles' incident
direction may vary for positrons and electrons. Even though no deviation of
their average livetimes is apparent (see section
\ref{section:fluxcalculation}), we account for this effect with a
second contribution to the systematic error of the positron
fraction. It is estimated from the mean variation of the difference in
livetime of positrons and electrons over the detector acceptance.
After propagation to the positron fraction, the systematic error due
to the East West Effect is well below 10~\% for all momentum bins,
except for the highest momenta above 26 GeV, where it amounts to
approximately 14~\% of the positron fraction value.

\section{Flux calculation}
\label{section:fluxcalculation}
As a crosscheck to the measurement of the positron fraction, presented
above, the absolute incident fluxes of electrons and positrons can be
calculated. The electron flux can then be compared to measurements by
other experiments and the results obtained previously by AMS-01.

One can calculate the differential flux for a given momentum bin $p$
of width $\Delta{}p$ from the measured
particle count $N(p,\theta,\phi)$ in this bin, the
detector acceptance $A(p,\theta,\phi)$, and the livetime
$T(p,\theta,\phi)$.
By the term {\it livetime}, we mean the effective amount of time
during which cosmic ray particles coming from outer space have the
opportunity to reach the detector. If -- as is the case with the
AMS-01 downward flux -- the livetime is only weakly depending on the
direction, the angular distribution of the particle count will follow
that of the acceptance. Then, the flux becomes
\begin{equation}
\diff{\Phi(p)}{p}=\frac{N(p)}{A(p)\cdot{}T(p)\cdot{}\Delta{}p} 
\end{equation}

While the detector acceptance is determined using a Monte Carlo method
based on the GEANT3 \cite{brun87} simulation of the detector, the live time is
calculated from an accurate approach of backtracing virtual particles
through the geomagnetic field.

\begin{figure}[!h]
\begin{center}
\begin{tabular}{cc}
\epsfig{file=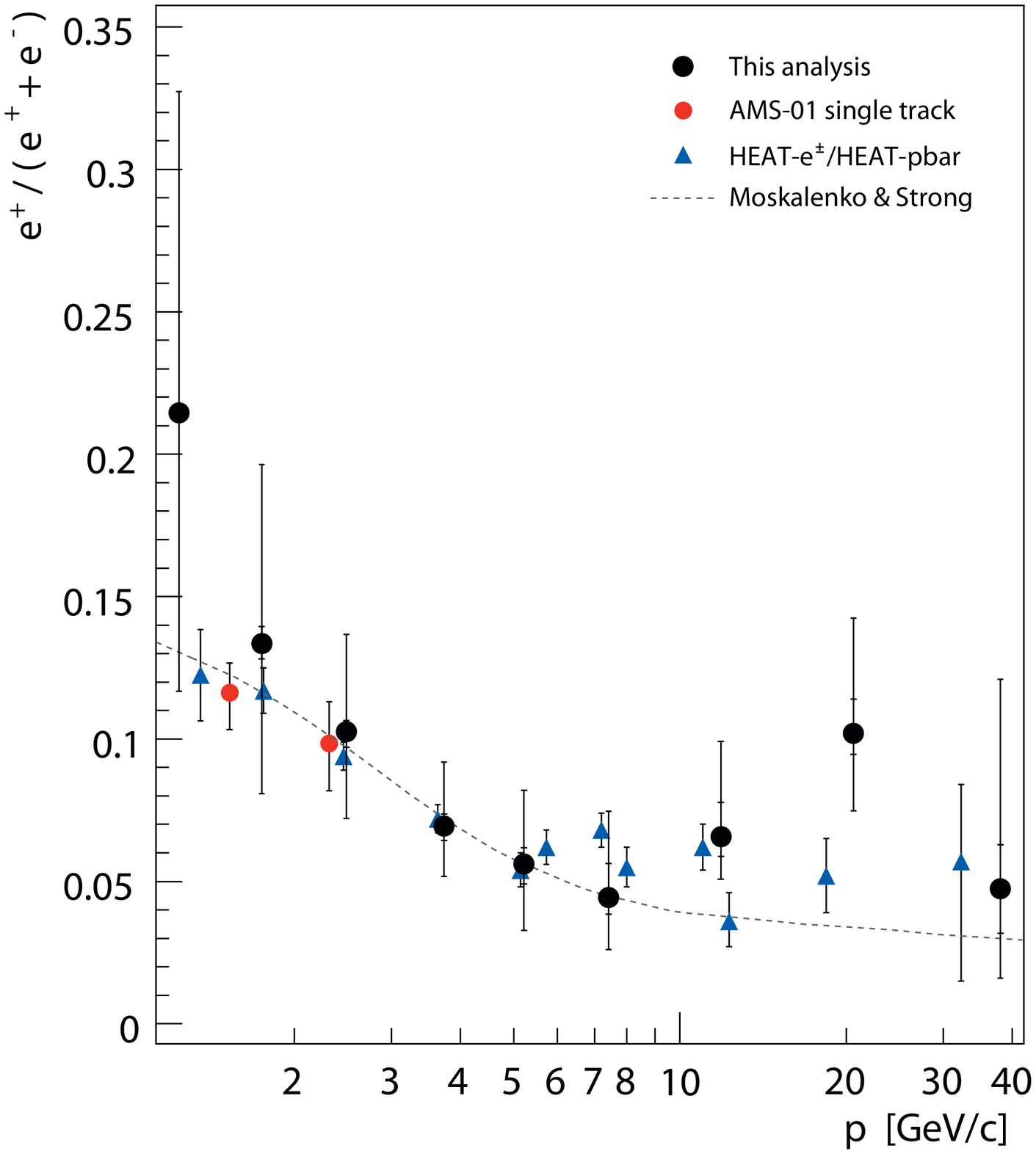,width=8cm}
&
\epsfig{file=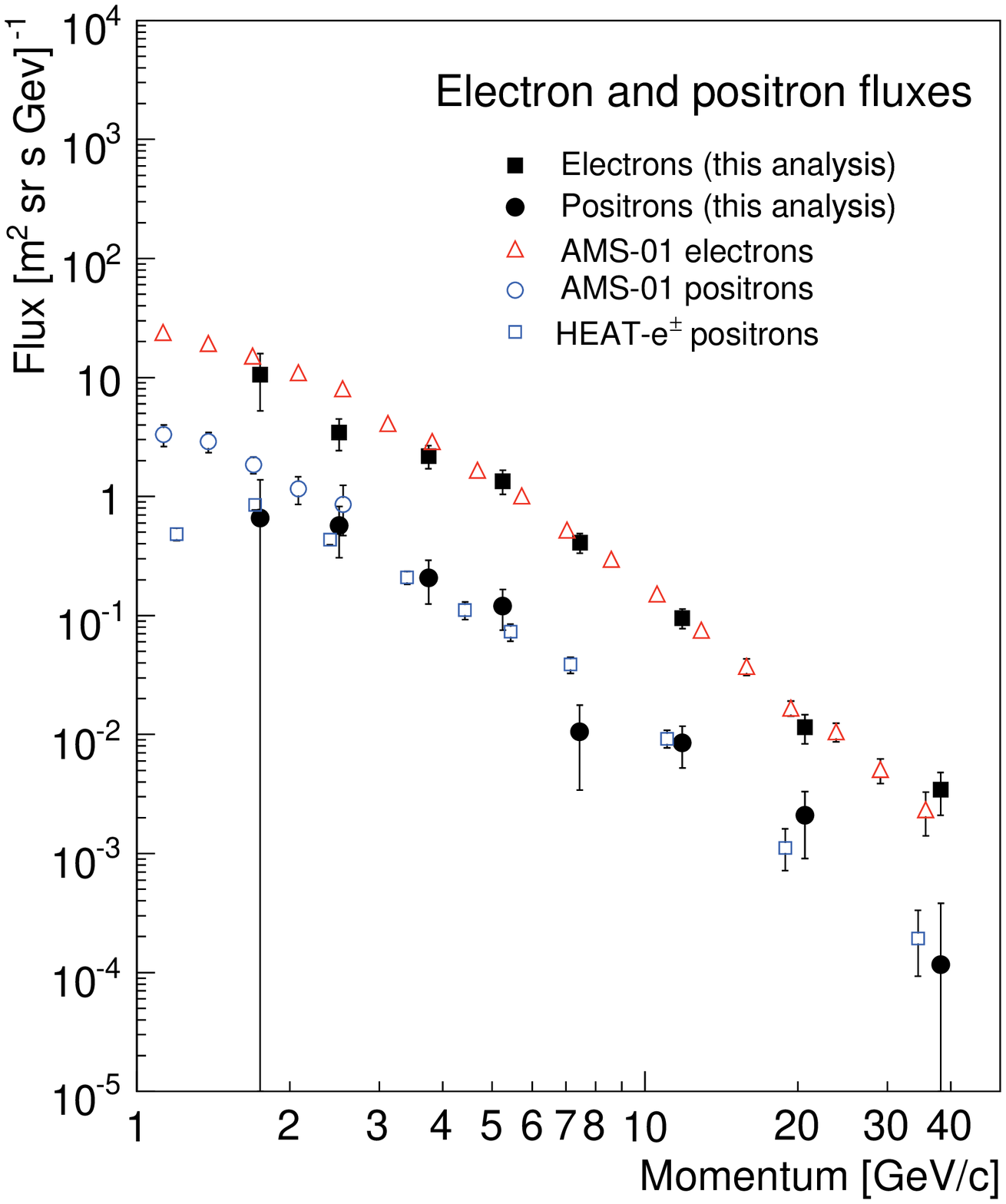,width=7.5cm}
\\
\end{tabular}
\caption{\label{fig:fraction}{\it Left:} The positron fraction $e^+/(e^{+}+e^-)$
measured in this analysis, compared with earlier results from
AMS-01 \protect\cite{alcaraz00a},
HEAT-$e^{\pm}$ \protect\cite{duvernois01a} and
HEAT-pbar \protect\cite{beatty04a}, together with a model calculation
for purely secondary positron production \protect\cite{moskalenko98a} (dashed line). The total error is given by the outer
error bars, while the inner bars represent the systematic contribution
to the total error.
\label{fig:fluxes}{\it Right: } The fluxes of downward going positrons and electrons measured
in this analysis, compared with earlier results from
AMS-01 \protect\cite{alcaraz00a} and HEAT-$e^{\pm}$ \protect\cite{duvernois01a}.
Error bars denote statistical errors only.}
\end{center} 
\end{figure}

In fig.~\ref{fig:fluxes} the fluxes of downward going positrons and
electrons, together with results published earlier by
AMS-01 \cite{alcaraz00a}, and HEAT-e$^{\pm}$ \cite{duvernois01a}, are
displayed with their statistical errors. The fluxes are in very good
agreement with previous measurements over the full momentum range,
except for a slight discrepancy in the electron fluxes between 2 and
3~GeV. Here, at low momentum in combination with low statistics, we
expect the inaccuracies of the backtracing through the geomagnetic
field to become the dominant source of systematic error to the
fluxes. However, for the positron fraction as a ratio of particle
counts, this effect widely cancels out.

\section{Outlook: AMS-02 aboard the International Space Station}
\begin{figure}[!h]
\begin{center}
\begin{tabular}{cc}
\epsfig{file=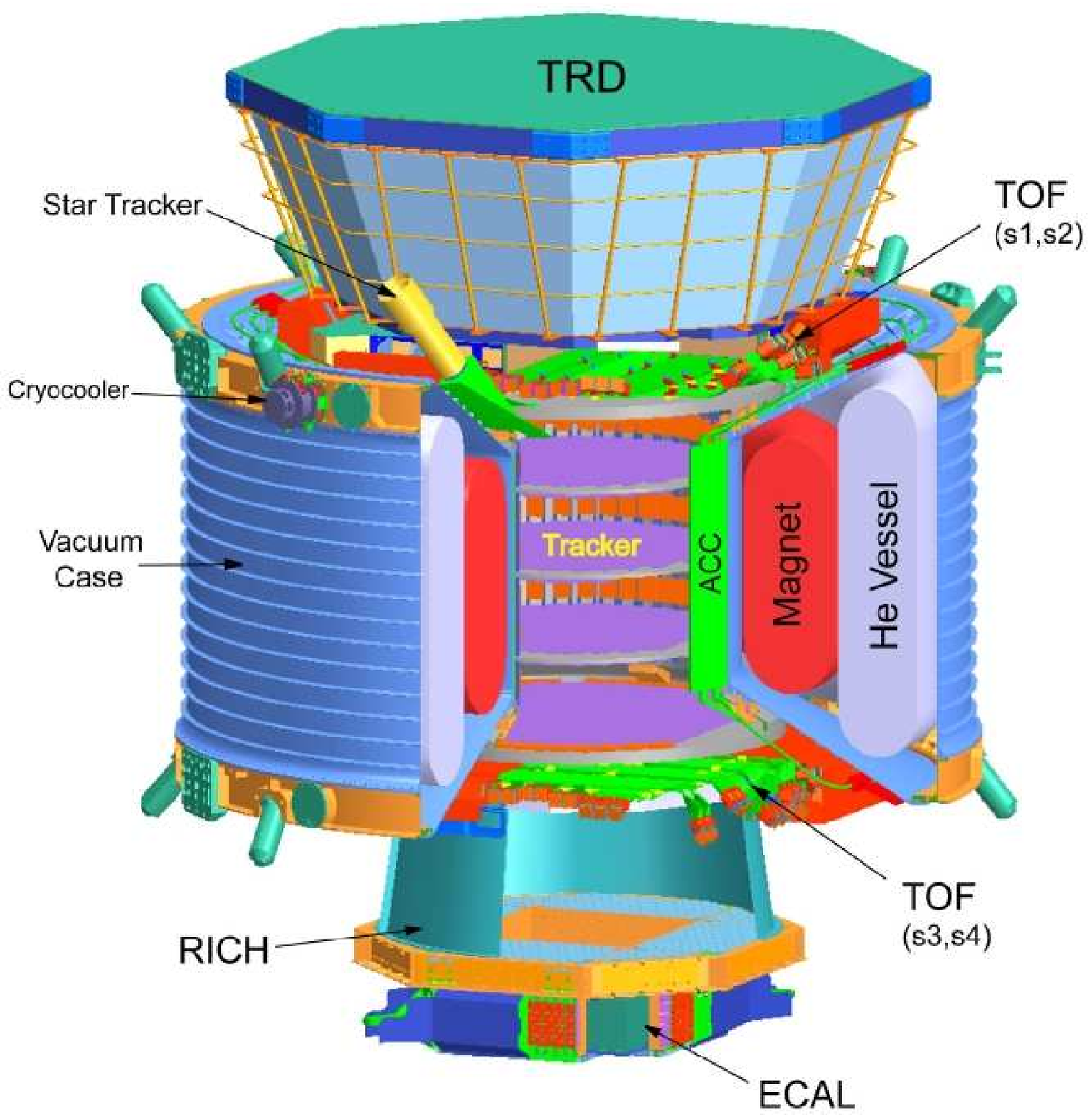,width=7cm}
&
\epsfig{file=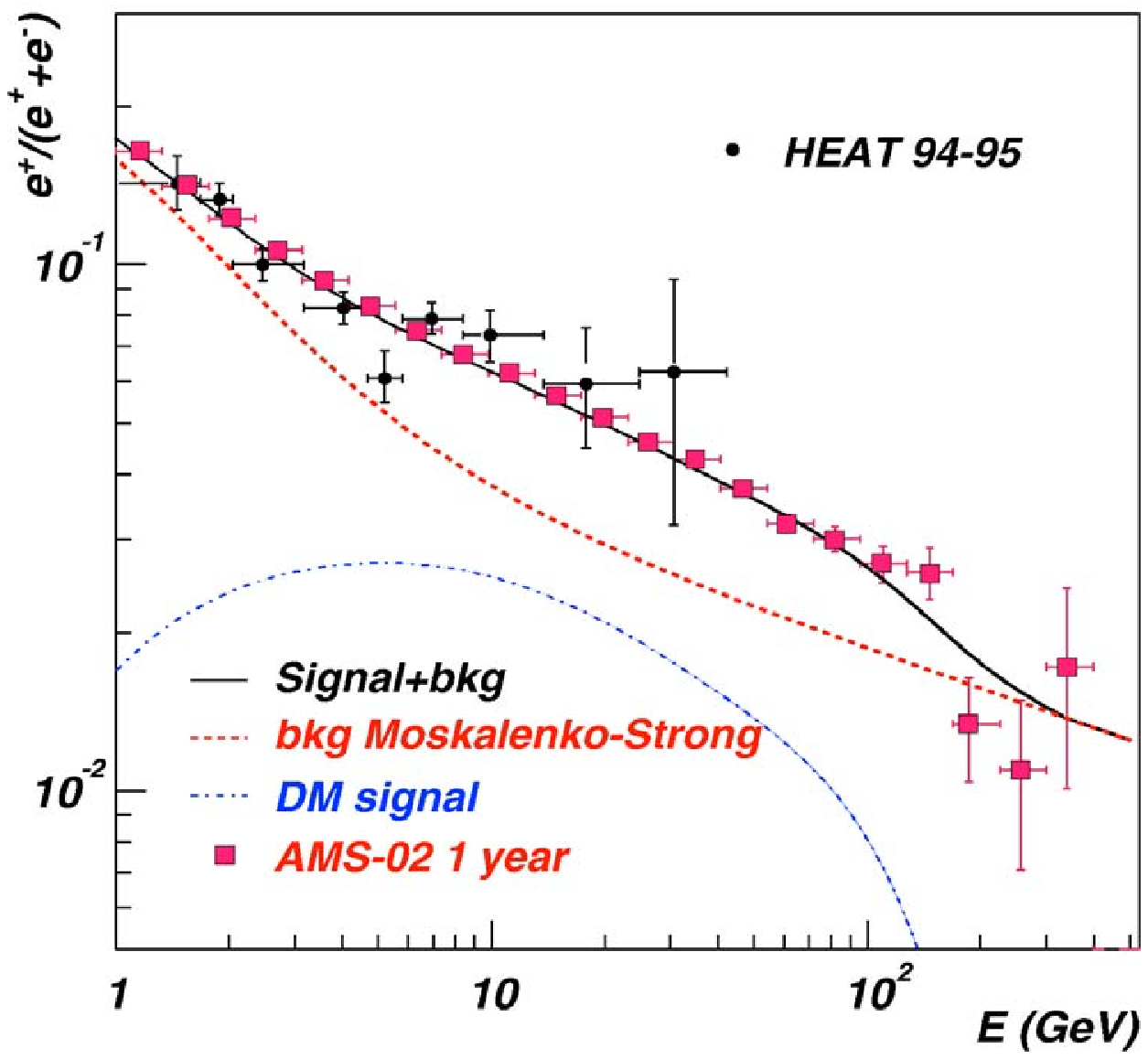,width=7cm}
\\
\end{tabular}
\caption{\label{fig:ams02parts}{\it Left:} View of the AMS-02
subdetectors.
\label{fig:ams02precision}{\it Right:} Statistical uncertainties on
positron fraction measurement with one year of AMS-02 data \protect\cite{bosio04a}.}
\end{center} 
\end{figure}

The successor to AMS-01, the AMS-02 experiment is currently under
construction by the collaboration \cite{leluc}. It will be installed on the
International Space Station (ISS) to conduct cosmic-ray spectroscopy
with unprecedentedly high precision. The subdetectors that will form
part of the experiment are depicted in figure \ref{fig:ams02parts} (left).\\
At its heart, a superconducting magnet of bending power
$BL^2=0.85$~Tm$^2$ forces charged particle onto a curved trajectory, which
is measured by the tracker. It consists of eight layers of
double-sided silicon microstrip sensors to allow rigidity
determination with an accuracy of $20\,\%$ at $500$~GV. The tracker
also has the vital task of finding the charge sign of a particle.\\
The
time-of-flight system, made of two double layers of scintillator
panels above and below the tracker, respectively, is used to measure
the velocity and charge of a particle and plays a main role in the
trigger system. The time resolution obtained is
$\Delta{}t=100$~ps.\\
On top of AMS-02, a transition radiation detector
(TRD), separates positrons from protons with a rejection power of
$10^2$ to $10^3$. To achieve this, it uses a radiator fleece and straw
drift tubes, filled with a Xe/CO$_2$ gas mixture, to detect the TR
photons.\\
The proton rejection will be enhanced by a factor of
$\mathcal{O}(1000)$ by an electromagnetic sampling calorimeter, made
of lead interleaved with scintillating fibres, which is mounted at the
bottom of the experiment.\\
A ring-imaging \v{C}erenkov counter for charge and velocity
measurements and an anti-coincidence counter for vetoing lateral
tracks complete the detector.

As an example for the capabilities of the AMS-02 detector, the
precision that can be obtained for the positron fraction is
demonstrated in figure \ref{fig:ams02precision} (right) for a one year
campaign. In the energy range
up to $100\,GeV$, the statistical error on the data points will be
completely negligible.\\
The main scientific goals of AMS-02 can be summarized as follows:
\begin{itemize}
\item Indirect search for dark matter in the positron, antiproton and
gamma spectra.
\item Search for cosmic antimatter ($\overline{\mathrm{He}}$ and
heavier nuclei).
\item Test propagation models for our Galaxy
($^{10}\mathrm{Be}/^9\mathrm{Be}$, $\mathrm{B}/\mathrm{C}$, and
others).
\end{itemize}

\section{Conclusions}
In this paper, we presented a new measurement of the cosmic ray positron
fraction up to energies of 50~GeV with the AMS-01 detector. Positrons
are identified by conversion of bremsstrahlung photons, an approach
which yields an overall proton rejection of more than $10^5$, and
allows to extend the energy range accessible to the experiment far
beyond its design limits, thus fully exhausting the detector's
capabilities. The results are consistent with those obtained in
previous experiments at large.

For the reconstruction of converted bremsstrahlung events, customized
algorithms for track finding and event reconstruction have to be
devised and implemented. We have shown that the background
is controllable and the overall uncertainty is dominated by the
statistical error due to the low overall cross section of the signal
process. 

Furthermore, the absolute lepton fluxes have been calculated and found
to match the earlier results. This required a new precise and
expensive lifetime calculation.

The discovery potential of AMS-02 has been briefly discussed.

\section*{Acknowledgements}
The authors wish to thank V. Choutko, MIT, and J. Alcaraz, CIEMAT, for
their vital help in various aspects of this work. The support of
J. Shin and G. Kim, Kyungpook National University, Korea, is gratefully
acknowledged.

\end{document}